\documentclass{pasj00}

\begin{document}
\SetRunningHead{H. Baba et~al.}{WZ Sge 2001 Outburst}
\Received{2001/01/01}
\Accepted{2001/01/01}

\title{Spiral Structure in WZ Sagittae around the 2001 Outburst Maximum}

\author{
  Hajime \textsc{Baba}\altaffilmark{1},
  Kozo \textsc{Sadakane}\altaffilmark{2},
  Yuji \textsc{Norimoto}\altaffilmark{3},
  Kazuya \textsc{Ayani}\altaffilmark{4},
  Masayuki \textsc{Ioroi}\altaffilmark{4}, \\
  Katsura \textsc{Matsumoto}\altaffilmark{5},
  Daisaku \textsc{Nogami}\altaffilmark{6},
  Makoto \textsc{Makita}\altaffilmark{7},
  Taichi \textsc{Kato}\altaffilmark{7}}

\altaffiltext{1}{Astronomical Data Analysis Center, National
  Astronomical Observatory of Japan, Mitaka, Tokyo 181-8588}
\email{hajime.baba@nao.ac.jp}
\altaffiltext{2}{Astronomical Institute, Osaka Kyoiku University,
  Asahigaoka, Kashiwara, Osaka 582-8582}
\altaffiltext{3}{Okayama Astrophysical Observatory, National
  Astronomical Observatory of Japan, Kamogata, Okayama 719-0232}
\altaffiltext{4}{Bisei Astronomical Observatory,
  Bisei, Okayama 714-1411}
\altaffiltext{5}{Graduate School of Natural Science and Technology, Okayama University, Okayama 700-8530}
\altaffiltext{6}{Hida Observatory, Kyoto University,
  Kamitakara, Gifu 506-1314}
\altaffiltext{7}{Department of Astronomy, Kyoto University,
  Sakyo-ku, Kyoto 606-8502}


\maketitle

\begin{abstract}
Intermediate resolution phase-resolved spectra of WZ Sge were
obtained on five consecutive nights (July 23 -- 27) covering the
initial stage of the 2001 superoutburst.
Double-peaked emission lines of He\textsc{II} at 4686 \AA,
which were absent on July 23, emerged on July 24
together with emission lines of C\textsc{III} / N\textsc{III}
Bowen blend.
Analyses of the He\textsc{II} emission lines using the Doppler
tomography revealed an asymmetric spiral structure
on the accretion disk.
This finding demonstrates that spiral shocks with a very short
orbital period can arise during the initial stage of an outburst
and may be present in all SU UMa stars.
\end{abstract}

\section{Introduction}

Dwarf novae are a subclass of cataclysmic variables (CVs), which
are close binary systems consisting of a white dwarf and a red
dwarf secondary, transferring matter via the Roche lobe overflow
(for a recent review, see \cite{war95}).  Dwarf novae show
frequent outburst, which are episodes of enhanced accretion
through the disk onto the central object.  These systems
provide good test cases for various theories of accretion disk
models.

WZ Sge is the prototype star of WZ Sge-type stars (originally
proposed by \cite{bai79}), which is a subclass of SU UMa-type
dwarf novae.  The remarkable outburst properties (a large
amplitude $\sim$ 8 mag., and the extremely long time interval
between outbursts about 33 years) show that WZ Sge is the most
extreme case among dwarf novae.  WZ Sge has a very short orbital
period of 1.37 h, and it is one of the few eclipsing dwarf novae,
whose inclination angle of the orbital plane is known to be 75
$\pm$ 2 degrees (for the detailed model of WZ Sge, see
\cite{sma93}).

One of the main long-standing problems concerning accretion disks
is the mechanism of angular momentum transport.  Spiral shocks
have long been proposed as a possible mechanism for the transport
(e.g. \cite{saw86}, \cite{spr87}, \cite{sav94}).
If the disk extends far enough in the Roche lobe, spiral shocks
are excited by the tidal field of the secondary and result in the
formation of two prominent spiral arms.  The gas in the disk loses
its angular momentum when it passes through the spiral shocks.

Analysis of the double-peaked emission lines using Doppler
tomography is a well-established method for imaging the accretion
disks in CVs (\cite{mar88}).  Emission lines seen in the spectra
arise from the accretion disk and can be seen in a double-peaked
shape as a result of Doppler motion of the disk within the binary.
The velocity of the material in the disk determines the line
profile.

Recently, spiral structures in the accretion disks of some dwarf
novae during outbursts have been discovered,
IP Peg (\cite{ste97}, \cite{har99}, \cite{mor00}),
EX Dra (\cite{joe00}), and U Gem (\cite{gro01}).
Their presence probed by the Doppler tomography method triggered a
renewed interest in spiral shock models.  The spiral pattern is
interpreted as evidence for shock waves, which is consistent with
the results of hydrodynamical simulations (e.g. \cite{ste99},
\cite{mak00}).

Aside from during outbursts, \citet{ski00} presented an extensive
set of Doppler maps of WZ Sge in quiescence using both optical and
infrared emission lines.  In these maps, the accretion disk
structure was found to be asymmetric and the bright spot region
was shown to be extended along the mass transfer stream, but these
structures didn't look like spirals.

The outburst of WZ Sge was discovered by an amateur astronomer on
2001 July 23.565 UT at a visual magnitude 9.7
(\cite{ish01iauc7669}), having occurred 10 years earlier than the
common expectation.  Receiving the report of the outburst, we
started time-series spectroscopic observations, to clarify the
evolution of the accretion disk structure at the very beginning of
the outburst.

\section{Observation}

The CCD spectroscopic observations were performed with the 91-cm
telescope at Okayama Astrophysical Observatory (OAO) and the
101-cm telescope at Bisei Astronomical Observatory (BAO) between
July 23 and 27.  According to VSNET
\footnote{\texttt{http://www.kusastro.kyoto-u.ac.jp/vsnet/}}, WZ
Sge was on the rising stage on July 23, and was at the maximum of
this superoutburst on July 24.

The intermediate dispersion spectra at the OAO covering 3950-5100
\AA\ with a resolution of 3000 were obtained using a CCD camera
(Andor DU440 camera with 2048 $\times$ 512 pixels Marconi CCD).
At the BAO, the low dispersion spectra covering 4250-6800 \AA\
were obtained using a peltier-cooled CCD camera (Mutoh CV-16II
with 1536 $\times$ 1024 pixles KAF-1600 CCD), while the high
dispersion spectra covering 6390-6750 \AA\ were obtained using a
liquid-nitrogen-cooled CCD camera (AstroCam 4200 series with
UV-coated 1024 $\times$ 256 pixels EEV CCD), resulting the
resolution of 1500 and 15000 respectively.  The journal of our
observations is summarized in Table \ref{tab:log}.

All of the raw frames were processed in the usual manner using IRAF.
The raw spectrum images were bias-subtracted and flat-fielded,
then stellar spectra were extracted from two dimensional spectrum
images.  Except for spectra taken on July 23, all spectra were
normalized to the continuum level, by means of the spline fitting.
The averaged signal to noise ratios (S/N) at the continuum level
were 60-100 for OAO data using exposure times of 180-300 s, and
25-40 for BAO data using exposure times of 120 s respectively,
which depend on the brightness of the object and the sky
condition.

Barycentric corrections to the observed times were applied before
the following analysis.  We adopted the binary ephemeris: $T_0
(BDJD) = 2437547.728868 + 0.05668784707 \times E$, where $T_0$ is
the mid-point of the eclipse, and a phase offset $\phi_c =
-0.022$, to convert the ephemeris to the inferior conjunction of
the binary.  Both of them are taken from \citet{ski00}.

\begin{table}
 \caption{Log of observations.}
 \label{tab:log}

 \begin{center}
 \begin{tabular}{rrrrr}
 \hline\hline
 UT (2001 July) & Orbit & No. of & Obs. & Observ- \\
 (start -- end) & covered & spectra & mode & atory \\
 \hline
 23.736-- 23.777 & --- &  6 & low  & BAO \\
 24.576-- 24.701 & 2.2 & 52 & ---  & OAO \\
 25.592-- 25.682 & 1.6 & 40 & ---  & OAO \\
 26.586-- 26.664 & 1.4 & 22 & ---  & OAO \\
 26.539-- 26.697 & 2.8 & 74 & high & BAO \\
 27.582-- 27.673 & 1.6 & 25 & ---  & OAO \\
 \hline
 \end{tabular}
 \end{center}
\end{table}

\section{Result}

\begin{figure}
  \begin{center}
    \FigureFile(88mm,60mm){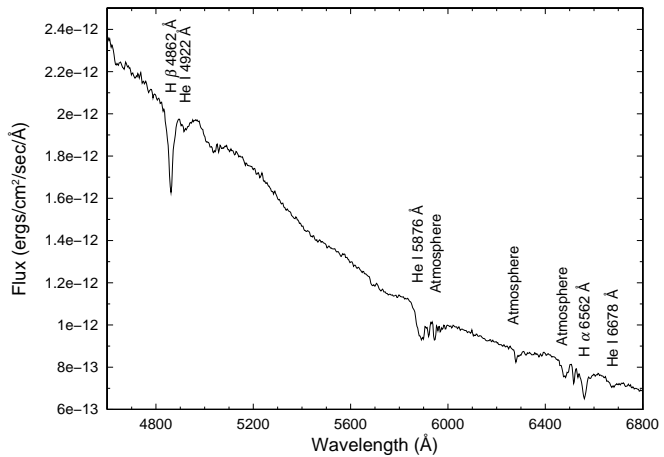}
  \end{center}

  \caption{The flux-calibrated spectrum on July 23,
  corresponding to the rising stage of the 2001 superoutburst of WZ Sge.
  Spectra taken on July 23 were corrected for instrumental
  response using the flux standard HD 8634.}
  \label{fig:bao723}
\end{figure}

Figure \ref{fig:bao723} is the flux-calibrated spectrum on July
23, which corresponds to the rising stage of the superoutburst.
It shows Balmer lines, H$\alpha$ and H$\beta$, in absorption on a
blue continuum.  He\textsc{I} lines (4922 \AA, 5876 \AA, 6678 \AA)
are also seen in absorption.  It is to be noted that no emission
or absorption component around the wavelength of He\textsc{II} and
C\textsc{III} / N\textsc{III} complex was observed on July 23.

\begin{figure}
  \begin{center}
    \FigureFile(88mm,60mm){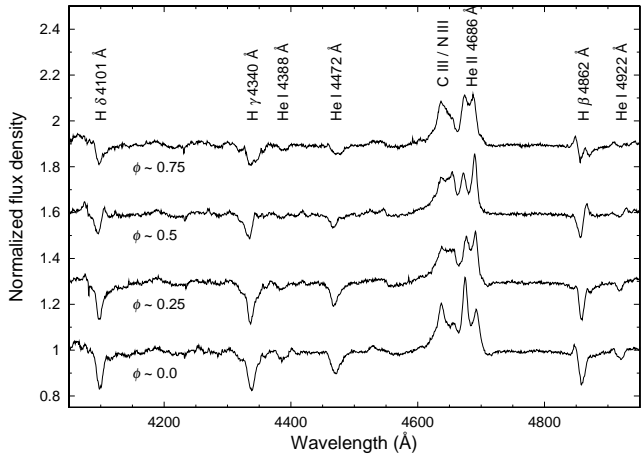}
  \end{center}

  \caption{Spectra of WZ Sge on July 24.
  The spectra are bound into four orbital phase slots.
  Before averaging, each spectra were shifted according to the
  radial velocity K${_1}$ sin($\phi$), with K$_{1}$ = 37 km s$^{-1}$
  (\cite{ste01}).
  An arbitrary offset were applied between spectra for display purpose.}
  \label{fig:oao724}
\end{figure}

Figure \ref{fig:oao724} shows the continuum-fitted spectra at four
different orbital phases, where the changes of the line profiles
are very impressive.  The most remarkable features in the spectra
on July 24 are the strong emission lines of the He\textsc{II}
4686 \AA\ and the C\textsc{III} / N\textsc{III} complex at 4640
\AA\ with double-peaked profiles, both of which were not observed
on July 23.  The peak intensities and the separation between peaks
vary with the orbital period.
H$\beta$ shows an emission component on the blue-wing between the
phase 0.7 -- 0.0, while it shows an emission component on the
red-wing between the phase 0.3 -- 0.5.  Nearly the same behavior
of H$\beta$ is observed on July 25, 26 and 27.  However, we notice
that no emission component of H$\beta$ was observed on July 23.
According to \citet{ste01}, who obtained spectroscopic data on
Aug. 6 and 13, prominent double-peaked emission components are
seen at both H$\alpha$ and H$\beta$, where the blue component is
much stronger than the red component.  Higher Balmer series
H$\gamma$ and H$\delta$ are observed as broad absorption
features between July 23 -- 27, together with neutral Helium lines
such as 4388 \AA, 4472 \AA\ and 4922 \AA.
The line feature resembles that obtained at the beginning of
the 1978 outburst reported in \citet{pat78} and in \citet{ort80},
both of which are taken at four days after the maximum. 

\begin{figure*}
  \begin{center}
    \FigureFile(40mm,60mm){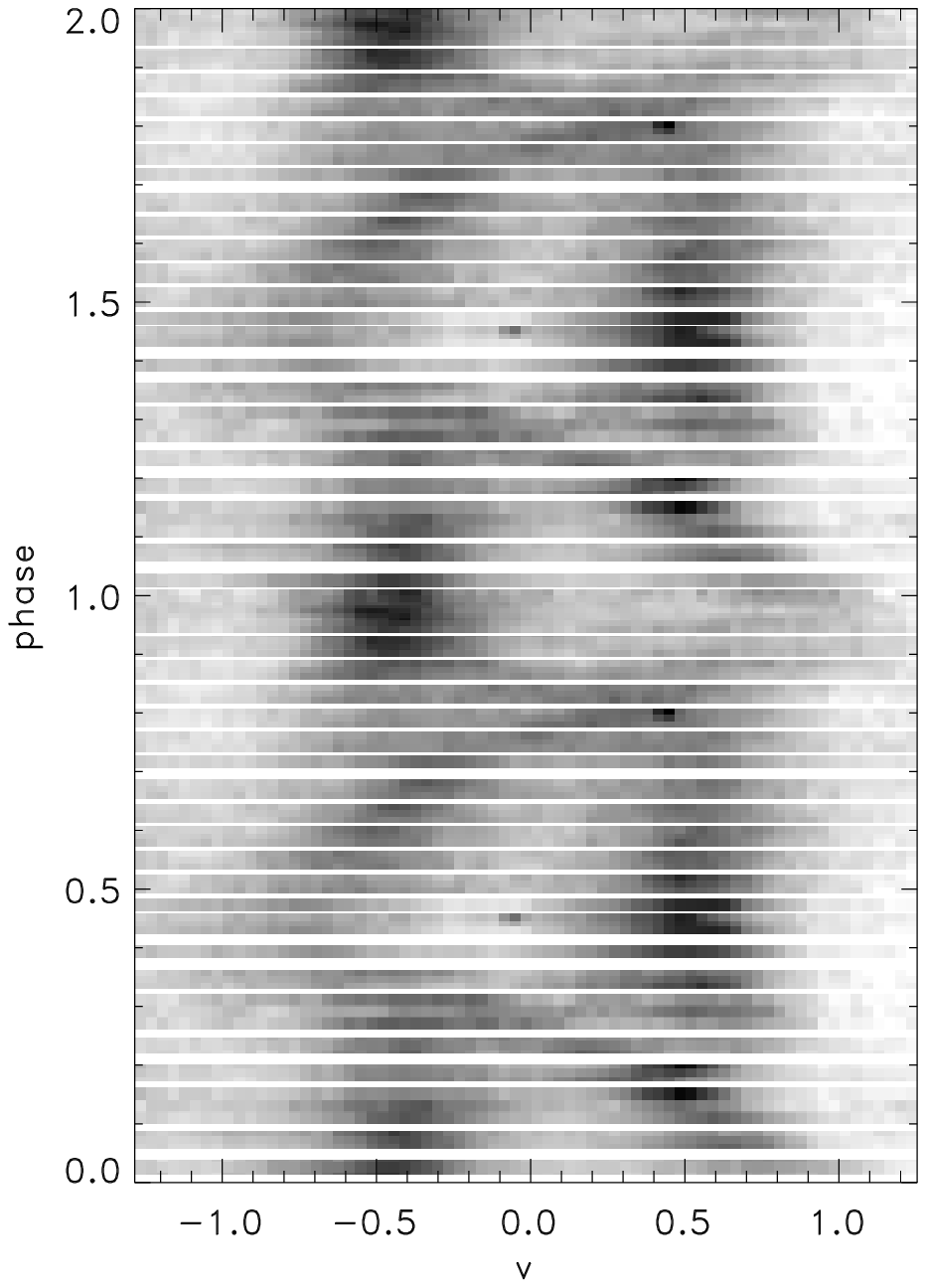}
    \FigureFile(40mm,60mm){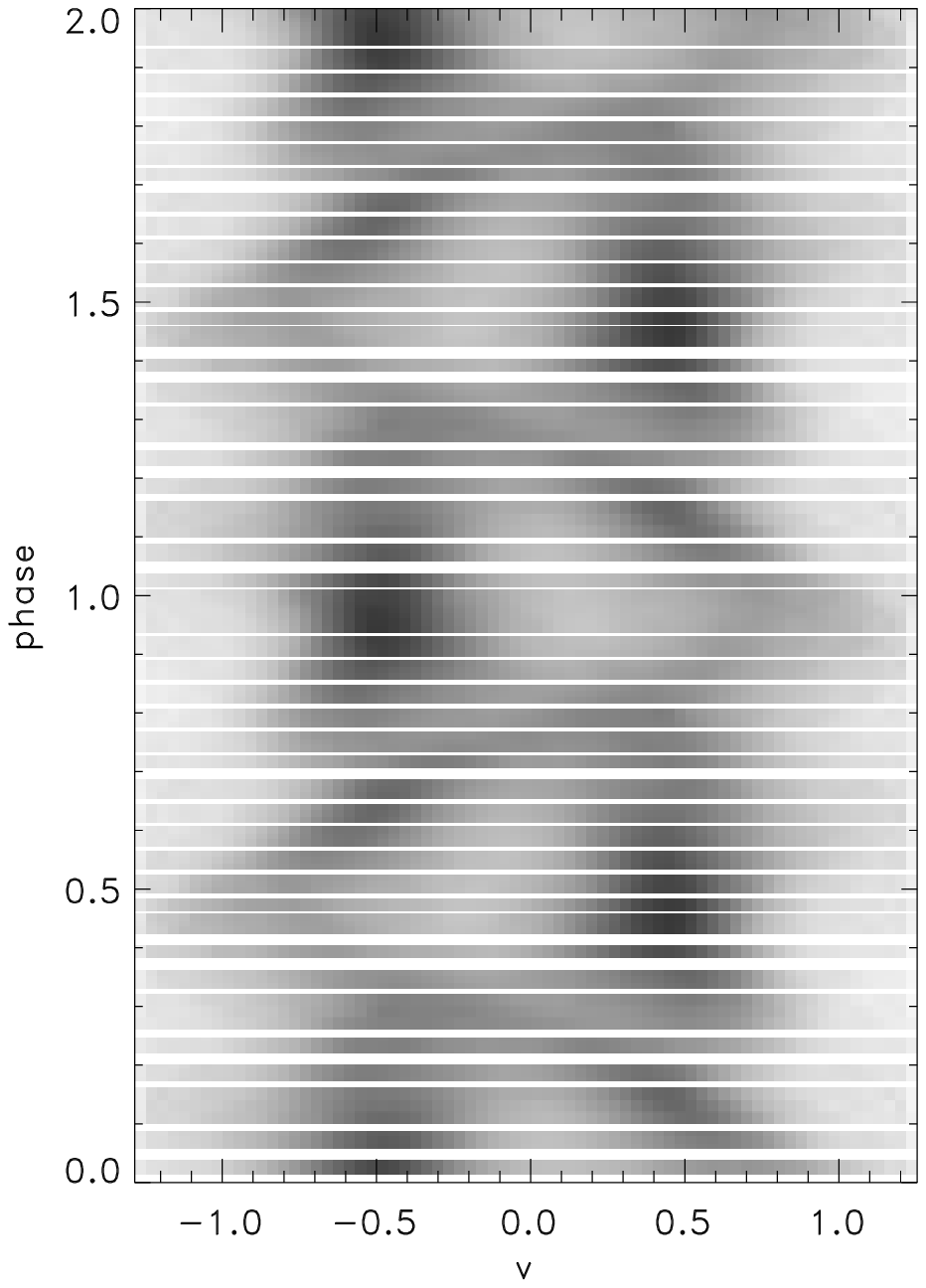}
    \FigureFile(60mm,60mm){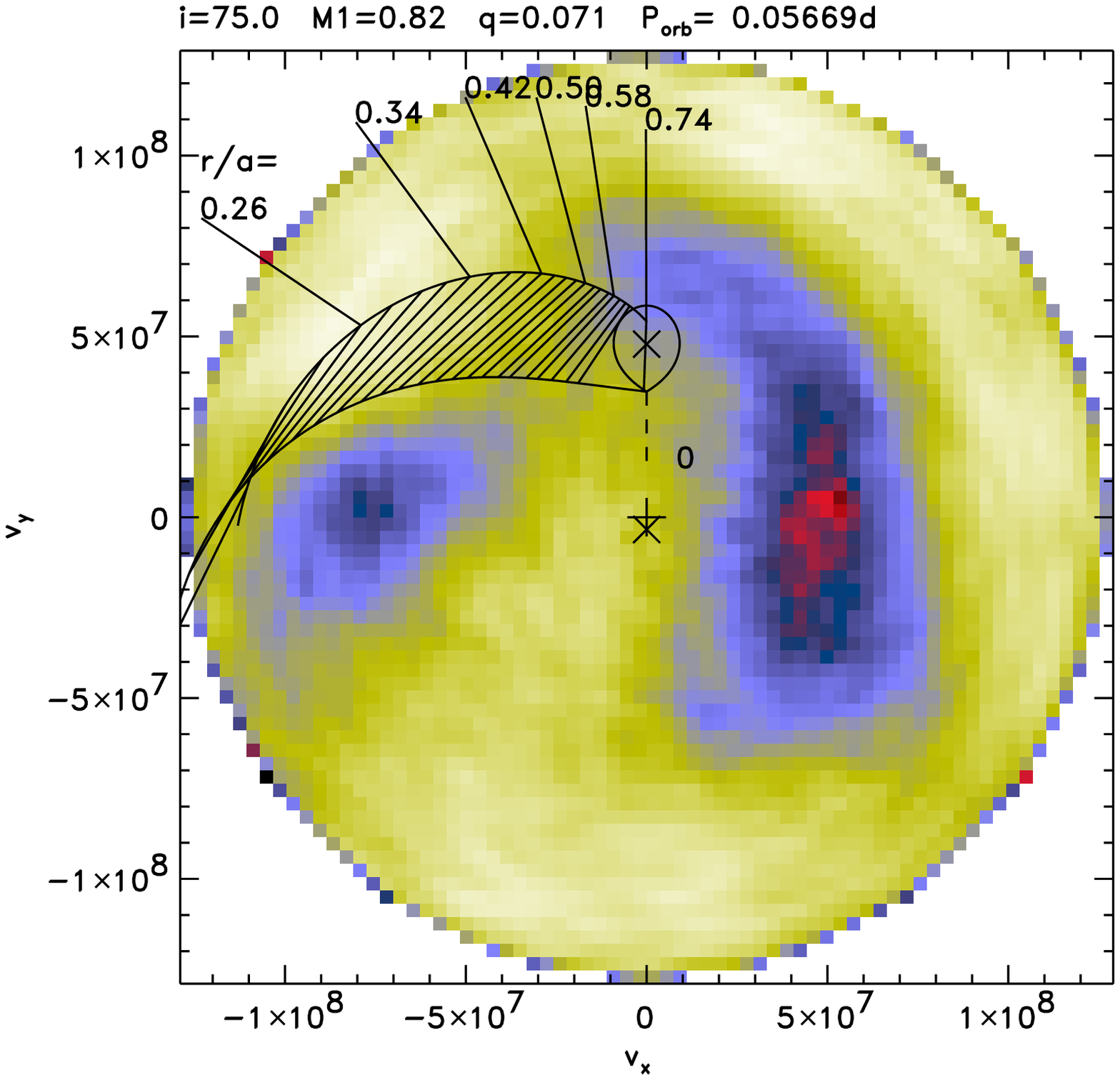} \\
   \vspace{0.5cm}
    \FigureFile(40mm,60mm){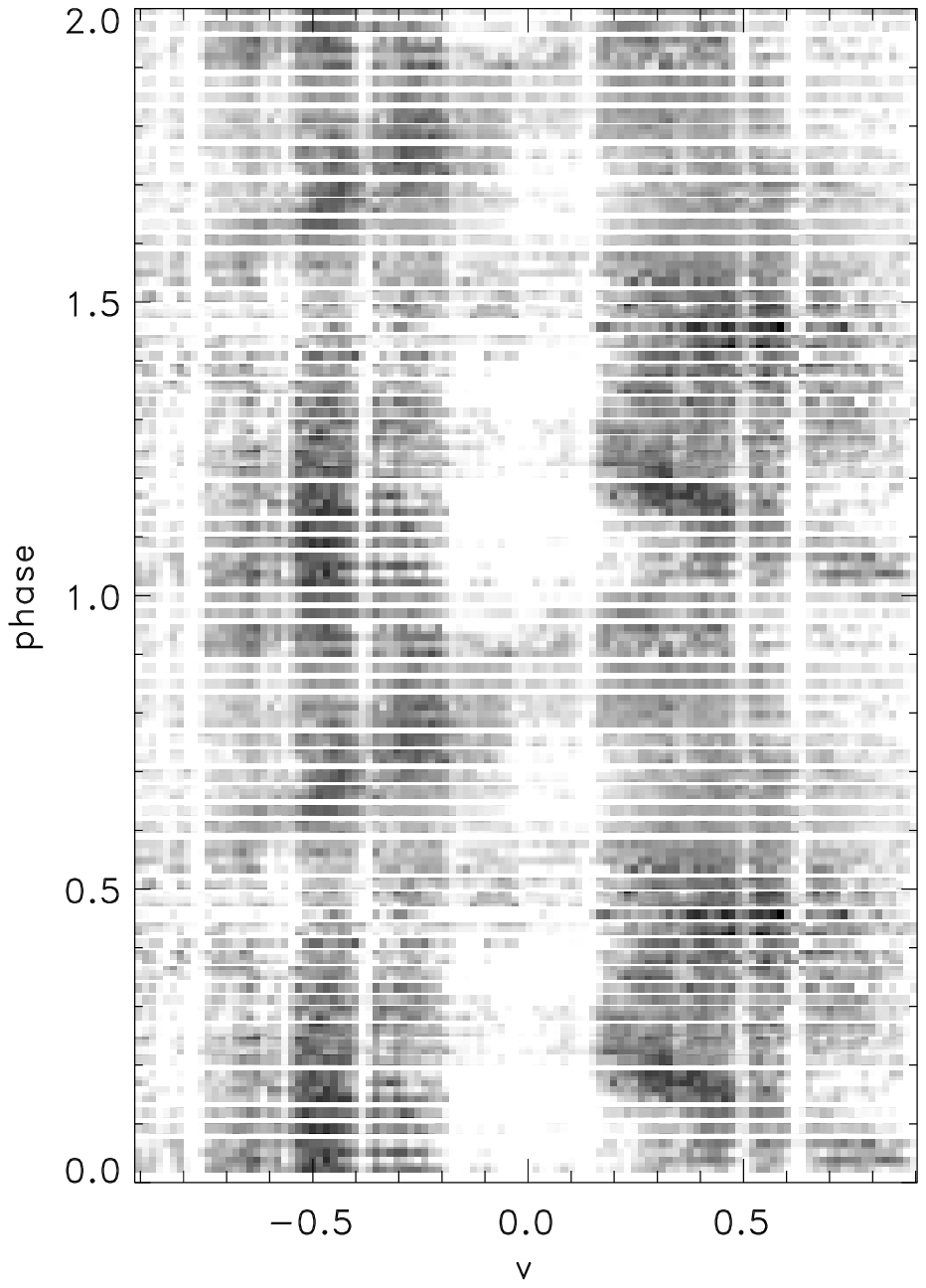}
    \FigureFile(40mm,60mm){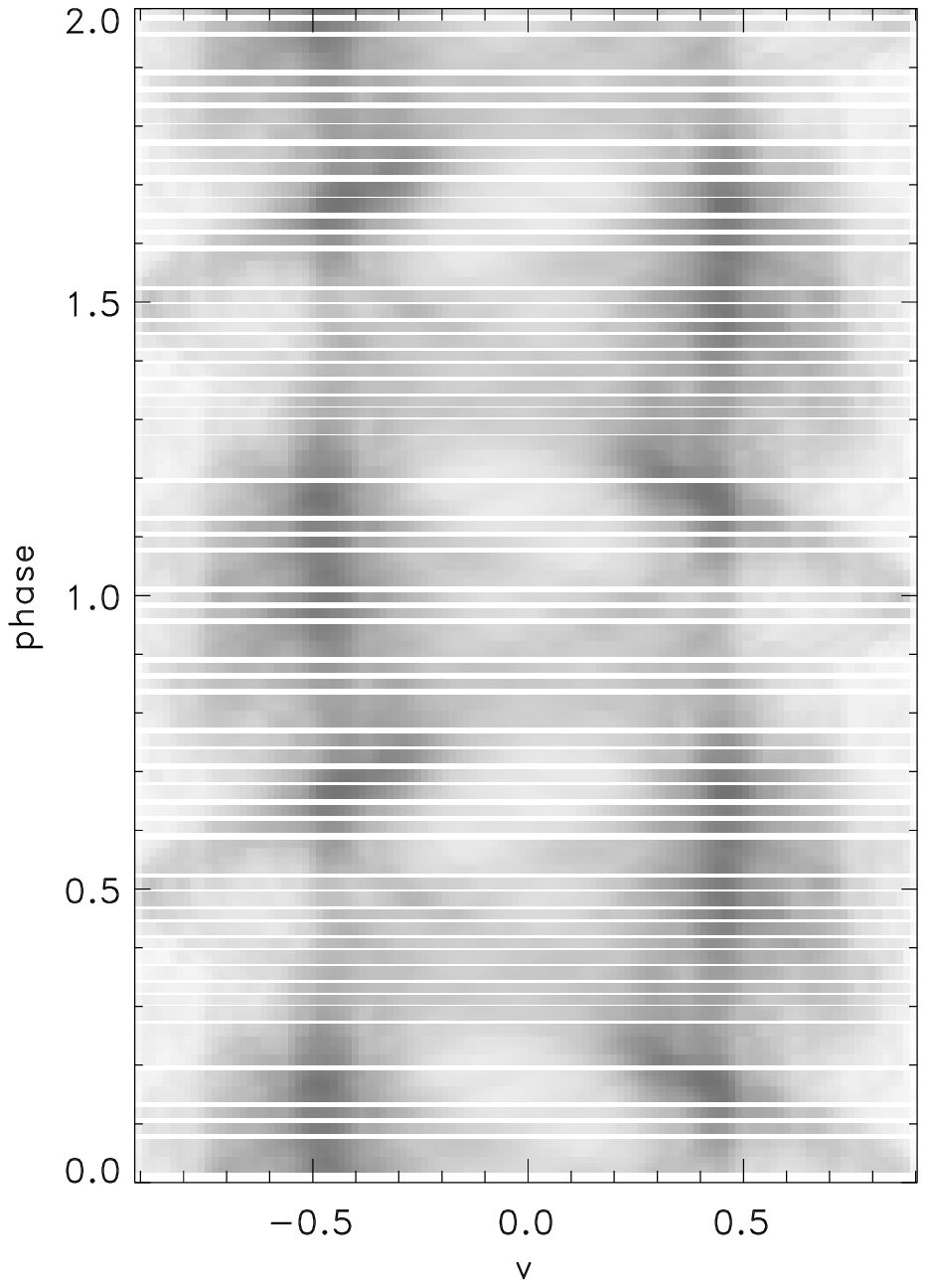}
    \FigureFile(60mm,60mm){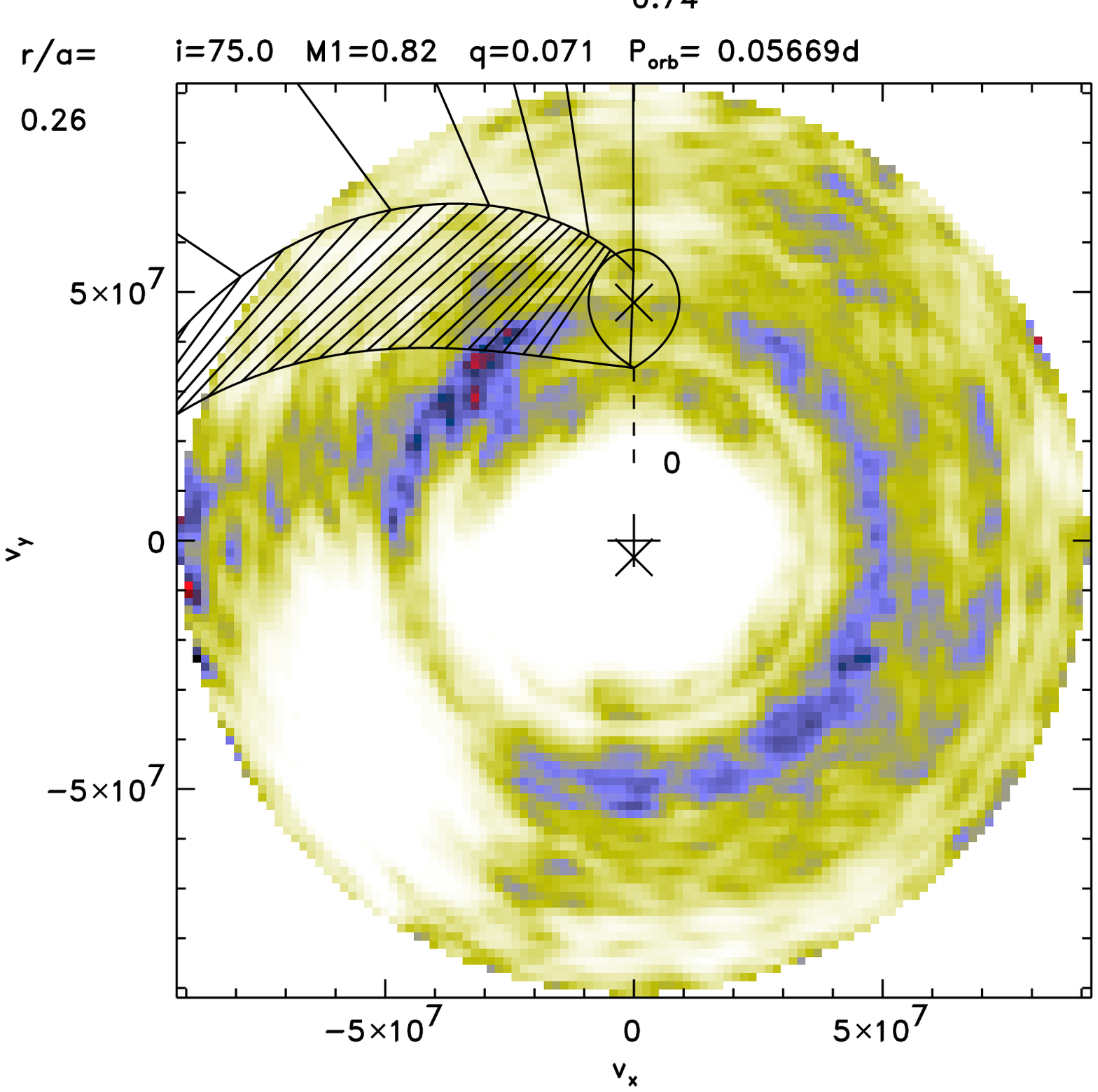}
  \end{center}

  \caption{Doppler maps of He\textsc{II} (top) and H$\alpha$
  (bottom).  Phase-folded spectra are shown in the left column,
  corresponding Doppler maps are in the right, and spectra
  reconstructed from the maps are shown in the middle column.
  
  Since the (partial) eclipse violates the Doppler mapping
  assumption of equal visibility at all phases, spectra obtained
  between -0.05 and 0.05 have been excluded from the
  reconstruction process.  However these parts of the data are
  included in the spectra shown in Fig. \ref{fig:dopmap} to
  clarify the change of the trails.

  The theoretical trajectory of the mass transferring stream are
  plotted in the Doppler images, together with the Keplerian
  velocity along the stream path.  Bars connecting the two arcs
  indicate correspondence in physical space, and are annotated
  with radius (r/a) and the azimuth relative to the primary.
  The system parameters including gamma velocity are taken from
  \citet{ski00}.}
  \label{fig:dopmap}
\end{figure*}

We constructed Doppler maps of WZ Sge using the time-resolved
spectra with the IDL-based fast-maximum entropy package developed
by \citet{spr98dopmap} \footnote{Available at
\texttt{http://www.mpa-garching.mpg.de/\%7Ehenk/}}.  Figure
\ref{fig:dopmap} shows the results for the strong emission line of
He\textsc{II} on July 24 and weak emission line of H$\alpha$ which
evolved on July 26.
The He\textsc{II} map displays the dominant accretion disk with
extended spiral arms, which constitutes an asymmetric spiral
structure.  The spiral arm in the right quadrants extends about
$\sim$ 135 degrees, and is stronger than the arm in the opposite
quadrants.  There is no observed evidence for the irradiation of
the secondary star nor for the hot-spot.  The images look quite
similar to the He\textsc{II} map of EX Dra (\cite{joe00}),
although the phase of the spirals seems to be different.
In constast, the H$\alpha$ image on July 26 significantly differs
from the He\textsc{II} maps.  It looks almost flat and there is no
remarkable feature such as the secondary star.

We clearly find an evidence for spiral structures in the accretion
disk at the beginning of the outburst of WZ Sge.
The Doppler maps of He\textsc{II} on July 25, 26 and 27 are almost
as same in strength and location as that on July 24.
\citet{ste01iauc7675} also reported that the He\textsc{II} and
C\textsc{III} emission lines were dominated by two-armed spiral
pattern on July 28.  Taking into account that no emission feature
of He\textsc{II} was detected on July 23 when the object was on
the rising stage, the shocks stood immediately just after the
maximum of the outburst, and persisted almost constant in strength
and location during the early stage of the outburst.

\section{Discussion}

The strong emmisivity of He\textsc{II} and almost no contibution
of H$\alpha$ indicate that the accretion disk had a high
temperature of a few tens of thousands K even at the edge region
at the early phase of this outburst.  In addition, no evidence of
the secondary in the HeII and Halpha indicates that the effect of
irradiation on the secondary was relatively weak.  This apparent
lack of strong irradiation on the secondary may have been a result
of the shielding of photons by a thickened accretion disk around
the superoutburst maximum, which is a natural consequence of an
extremely hot (a few tens of thousands K) disk.

All of the three systems in which spiral arms have been observed,
i.e. IP Peg, EX Dra and U Gem, are the systems above the period
gap.  \citet{har99} suggested the possibility that the spiral
shocks only develop in systems above the period gap, since higher
mass-ratio binaries are expected to have a heavier secondary star
which induces stronger tidal torques.  Nevertheless, WZ Sge is the
system below the period gap, together with other SU UMa-type
stars.
The present discovery suggests the possibility of the existence of
spirals in other SU UMa-type stars during outburst, although there
has been a report of nagative detection (e.g. OY Car in outburst
(\cite{har96})).  There may be
a selection effect since the outburst of the short period systems
are difficult to expect, and as a result, phase-resolved
spectroscopy of the short period systems at the very beginning of
outburst must be difficult.  Further strategic observations are
strongly encourged to confirm the speculation.

According to \citet{ish01iauc7669}, the humps with the orbital
period called ``early superhumps'' were observed at the very early
stage of this superoutburst, which is the characteristics of WZ
Sge-type stars.  Even though the origin of ``early superhumps'' is
still unknown, the immediate evolution of spirals suggests that 
the double peak of ``early superhumps'' may be understood as a
reflection of the two arms of the spiral structure.
If so, the extended wing in the right half region of He\textsc{II}
Doppler map corresponds to the secondary maxima of the early
superhumps ($\phi \sim 0.2$), and the weak component in left
quadrant of He\textsc{II} Doppler map corresponds to the primary
maxima of the early superhumps ($\phi \sim 0.6-0.7$).
This intensity inversion is left to be a problem.

Although there is no observation of periodic modulation like
``early superhumps'' in the systems where spiral structures were
detected, it may be possible that different mechanisms drive
two-armed spiral patterns during outburst in WZ Sge and in some
dwarf novae above the period gap.
In a theoretical work, \citet{lin79} suggested that only
an accretion disk in a small mass-ratio system (such as $q<0.1$)
could show a strong spiral dissipation pattern with the 2:1
Lindblad resonance.
\citet{odo90} pointed out that complex oscillations observed in AM
CVn stars with more extremely low mass-ratios than WZ Sge may be
understandable with multiple standing shocks demonstrated by
\citet{lin86}.  We suppose that such mechanisms of standing shock
of AM CVn stars may be applicable in the case of WZ Sge.

\bigskip

We are grateful to Tomohito Ohshima for his detection and early
notification of the long-awaited superoutburst.  We are also
grateful to Prof. Masanori Iye, director of OAO, who permitted us
promptly to carried out the Time-Of-Opportunity observations using
the OAO 91-cm telescope.  We wish to thank Dr. H. C. Spruit for the
use of his DOPMAP programs.
We acknowledge Prof. Yoji Osaki for the useful discussion.

\end{document}